\documentclass[
    aps,
    pra,
    reprint,
    amsmath,
    amssymb,
    longbibliography,
    superscriptaddress
]{revtex4-2}

\usepackage{graphicx}
\usepackage{bm}
\usepackage{physics}
\usepackage{dcolumn}
\usepackage{hyperref}
\usepackage{xcolor}
\usepackage{amsmath}
\usepackage{mathrsfs}

\DeclareMathOperator{\diag}{diag}

\begin{document}


\title{CCZ gate beyond the strong Rydberg-blockade regime}


\author{Snigdhadev Ray}
\email{snigdhadevray@gmail.com}
\affiliation{
School of Physical Sciences,
Indian Association for the Cultivation of Science,
Kolkata 700032, India
}

\author{Sinchan Snigdha Rej}
\email[Contact author: ]{sinchansrej@gmail.com}
\affiliation{
School of Physical Sciences,
Indian Association for the Cultivation of Science,
Kolkata 700032, India
}

\date{\today}


\begin{abstract}

We present a protocol for implementing a controlled-controlled-Z (CCZ) gate on a neutral-atom platform, employing three individually addressable atoms confined in a linear array of optical tweezers and operating under finite Rydberg-Rydberg interaction (RRI). The protocol combines successive excitation of the Rydberg atoms with analytically determined pulses and uses two Rydberg levels to access opposite-sign interaction shifts. We show two different paths to get the conditional phase in the target atom: a pair of target pulses with opposite phases and a single detuned target pulse followed by a local phase correction. We derive analytical expressions for the time-integrated Rydberg populations to calculate the associated decay errors and also evaluate gate performance through master-equation simulations. Numerical comparisons with a fast Rydberg-antiblockade (RAB)-based CCZ protocol show improved gate performance at larger interatomic separations within the parameter regime considered. Monte Carlo sampling further quantifies the sensitivity to thermally induced position fluctuations. These results identify the tradeoffs between gate duration, Rydberg-state exposure, and interaction sensitivity, providing an analytically specified route to native three-qubit phase gates beyond the strong-blockade regime.

\end{abstract}



\maketitle


\section{Introduction}

Neutral atoms confined in optical-tweezer arrays provide a platform that integrates high-fidelity coherent qubit manipulation with strong, tunable Rydberg-mediated interactions~\cite{SaffmanWalkerMolmer2010,Adams2019,Saffman2016JPB}. This combination enables programmable architectures for quantum computation and quantum simulation~\cite{Morgado2021,Wu2021ChinPhysB,Browaeys2020}, building on foundational proposals for Rydberg two-qubit gates~\cite{Jaksch2000,Lukin2001} and for analog simulation of spin Hamiltonians~\cite{Weimer2010}. Initial experimental demonstrations of Rydberg blockade between individually trapped atoms~\cite{Urban2009,Gaetan2009} were subsequently leveraged to generate Bell states and to implement controlled-NOT operations~\cite{Wilk2010,Isenhower2010,Zhang2010}. Later work generalized entangling protocols to two-dimensional tweezer arrays~\cite{Maller2015,Graham2019} and achieved substantial improvements in gate performance across both alkali and alkaline-earth(-like) atomic species~\cite{Levine2018,Madjarov2020,Evered2023}. More recent milestones—including algorithmic demonstrations~\cite{Graham2022}, coherent transport of atoms within reconfigurable arrays~\cite{Bluvstein2022}, and logical-level processing~\cite{Bluvstein2024}—have intensified interest in native gate sets optimized for circuit efficiency and for the requirements of quantum error correction~\cite{Cong2022}.
Native multiqubit gates can reduce the entangling-gate count relative to decompositions into elementary two-qubit operations.
Rydberg implementations include multibit controlled-NOT gates~\cite{Isenhower2011,Rej_2026}, the Deutsch--Toffoli gate family~\cite{Shi2018}, parity-controlled gate~\cite{Rejpsc_2026} and experimentally demonstrated three-qubit operations~\cite{Levine2019}.
Related approaches exploit three-body F\"orster resonances~\cite{Beterov2018}, conditional driving in Ising-coupled systems~\cite{Rasmussen2020}, or adiabatic evolution in interacting excited-state manifolds~\cite{Khazali2020}.
Asymmetric interactions provide further routes to collective entanglement and multiqubit phase gates~\cite{SaffmanMolmer2009,Wu2010PRA,Young2021}.
Here we focus on the CCZ gate, which changes the sign of $\ket{111}$ while leaving the other computational basis states unchanged.
Hadamard rotations on the target convert it into a Toffoli gate.

Conventional blockade gates suppress simultaneous Rydberg excitation by requiring interaction shifts $|V|\gg\Omega$, with $V$ and the driving Rabi frequency $\Omega$ expressed in angular-frequency units.
At finite blockade strength, off-resonant coupling to doubly excited states produces residual population and unwanted light shifts; for standard square-pulse sequences, the leakage probability is typically of order $(\Omega/V)^2$.
Reducing $\Omega$ suppresses this leakage but prolongs the time spent in Rydberg states, increasing decay errors, whereas faster excitation increases the sensitivity to imperfect blockade~\cite{SaffmanWalker2005,ZhangGill2012}.
Since van der Waals (vdW) interactions scale as $V\propto l^{-6}$, maintaining a fixed blockade ratio as the separation $l$ increases requires slower driving and therefore greater decay exposure.
Finite confinement introduces an additional source of error through fluctuations of the interatomic separation.
For small positional deviations $\delta l$ at fixed vdW coefficient, $\delta V/V\simeq-6\delta l/l$; these interaction variations modify leakage and conditional phases, and can prevent interaction-calibrated excitation cycles from closing~\cite{Shi2017BlockadeFree,ShiLu2021}.
Relaxing the strong-blockade condition therefore requires explicit attention to both decay exposure and sensitivity to interaction fluctuations.
Several two-qubit proposals explicitly incorporate finite interaction shifts into the desired evolution.
Shi~\cite{Shi2017BlockadeFree} uses carefully chosen pulses to return the population to the computational states even when \(|V|\sim\Omega\). This removes blockade error when the interaction strength and pulse parameters match the design. The van der Waals scheme uses both positive and negative interaction shifts.
Shi and Lu~\cite{ShiLu2021} use a phase reversal of the target drive to cancel the unshifted evolution, while interaction-detuned Rabi cycles accumulate a tunable conditional phase using a single Rydberg level.
This permits entanglement between distant atoms with weak vdW interactions.
Cole \emph{et al.}~\cite{Cole2026} introduce a detuned target pulse into a $\pi$--$2\pi$--$\pi$ sequence, closing both conditional trajectories without strong blockade.
Faster control-atom pulses reduce decay exposure, while optimized target-pulse phases improve robustness against static variations of the interaction strength or Rabi frequency at the cost of longer operation times.
These studies distinguish cancellation of the ideal blockade error from robustness to imperfectly known interactions. Recent proposals address extended-range operation through several distinct mechanisms.
Bergonzoni \emph{et al.}~\cite{Bergonzoni2026} use resonant dipole-dipole spin exchange, scaling as $l^{-3}$, together with optimally shaped laser pulses to realize two-qubit iSWAP gates between atoms separated by more than $20\,\mu\mathrm{m}$.
Their approach combines excitation, exchange, and deexcitation within a continuous control pulse and accounts for residual vdW interactions and motional effects. A complementary route connects distant qubits through auxiliary atoms.
Doultsinos and Petrosyan~\cite{Doultsinos2025} propose a CZ gate mediated by a chain whose Rydberg excitations are adiabatically transferred into an antiferromagnetic configuration. Chirped excitation and deexcitation pulses, together with reversal of the interaction sign, cancel the dynamical phase while retaining a conditional geometric phase. The broader landscape of Rydberg gates also includes shaped pulses, symmetric adiabatic driving, and optimal control~\cite{Theis2016,Saffman2020,Jandura2022}; ion-mediated interaction \cite{PhysRevA.110.062618}; Rydberg dressing and weak-coupling schemes~\cite{Jau2016,Jo2020}; and conditional operations based on electromagnetically induced transparency and dark-state dynamics~\cite{Muller2009,Petrosyan2017,Rao2014, Rej_2026}.
Geometric and holonomic protocols exploit cyclic evolution \cite{Rejnjp_2026}, including schemes accelerated by shortcuts to adiabaticity~\cite{Moller2008,Liao2019,Zhao2017,Kang2018}, while Stark-tuned F\"orster resonances provide control of the interaction channels~\cite{Beterov2016}.
Antiblockade compensates interaction shifts with laser detunings~\cite{Ates2007,Amthor2010} and supports dissipative entanglement preparation~\cite{Carr2013} and conditional gate sequences~\cite{Su2017,Su2020EPL}.For a native three-qubit operation, Ashkarin \emph{et al.}~\cite{Ashkarin2025} propose CC$\Phi$ gates using radio-frequency-induced three-body F\"orster resonances in a linear array of $^{87}\mathrm{Rb}$ atoms.
Their scheme provides a tunable conditional phase through coherent transfer between collective Rydberg configurations, with calculations for neighboring atoms separated by $10\,\mu\mathrm{m}$.
It requires control of the radio-frequency field and the dc electric field that selects the resonant channel. Delakouras \emph{et al.}~\cite{Delakouras2026} extend the principle of \cite{Doultsinos2025} to native $C_kZ$ gates in star-graph arrangements driven by global pulses, and to distant multiqubit gates connected by atomic buses.
The phase depends on the parity of the intermediate Rydberg-excitation pattern.

In this work, we propose a six-step CCZ gate protocol on a neutral atom architecture where three individually addressable atoms are trapped in a linear array of equidistant optical tweezers, with the RRI magnitude comparable to the relevant Rabi frequencies.
 The protocol integrates sequential excitation of Rydberg atoms using analytically derived pulse sequences and exploits two distinct Rydberg states to access interaction-induced energy shifts of opposite sign.
We construct two alternative paths for the target-atom operation: a pair of pulses with opposite Rabi amplitudes, and a single detuned pulse followed by a local phase correction.
Both paths are designed to return the population to the computational subspace and produce the same conditional $\pi$ phase.
  The physical resources differ from those of the extended-range proposals above.
Our conditional phase uses vdW shifts rather than the resonant exchange of Ref.~\cite{Bergonzoni2026} or the rf-induced three-body transfer of Ref.~\cite{Ashkarin2025}.
All three atoms encode computational qubits, so no additional atomic bus is introduced.
This construction requires individual addressing, suitable Rydberg levels, and calibration of the interaction-dependent pulse parameters.
It relaxes the strong-blockade requirement; the accessible separation therefore remains constrained by the gate duration and Rydberg lifetimes. We quantify these constraints by deriving the time-integrated Rydberg populations for both paths and evaluating their decay errors.
Master-equation simulations characterize the gate fidelity as a function of operation time and interatomic separation, including a comparison with the fast antiblockade CCZ protocol of Ref.~\cite{Su2017} under the parameters considered here.
Monte Carlo sampling of atomic positions assesses sensitivity to thermal position uncertainty.
Collectively, these analyses elucidate the trade-offs among gate duration, Rydberg-state exposure, and interaction sensitivity within the finite-blockade regime.

Section~\ref{sec:model} introduces the protocol.
Section~\ref{err} derives the Rydberg-decay contribution to the gate error, and Sec.~\ref{sec:results} presents the numerical fidelity and position-fluctuation analyses.
Section~\ref{sec:conclusion} concludes the paper.
\begin{figure}
    \centering
    \includegraphics[width=\linewidth]{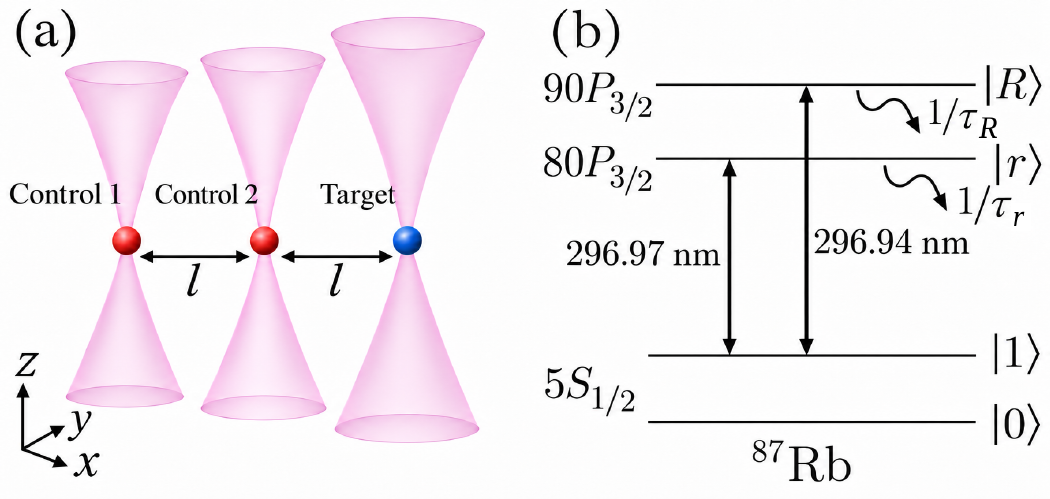}
    \caption{(a) Two control atoms (red ones) and the target atom (blue one) are trapped in three different equidistant optical tweezers. The interatomic spacings are defined by $l$. (b)Atomic level scheme for $^{87}\mathrm{Rb}$. The qubit states $\ket{0}$ and $\ket{1}$ are encoded in the $F=1$ and $F=2$ hyperfine levels of the $5S_{1/2}$ ground state, respectively. The Rydberg states $\ket{r}$ and $\ket{R}$ correspond to $80P_{3/2}$ and $90P_{3/2}$, respectively.}
    \label{fig1}
\end{figure}

\section{Theoretical model}
\label{sec:model}

\begin{figure*}
    \centering
    \includegraphics[width=\textwidth]{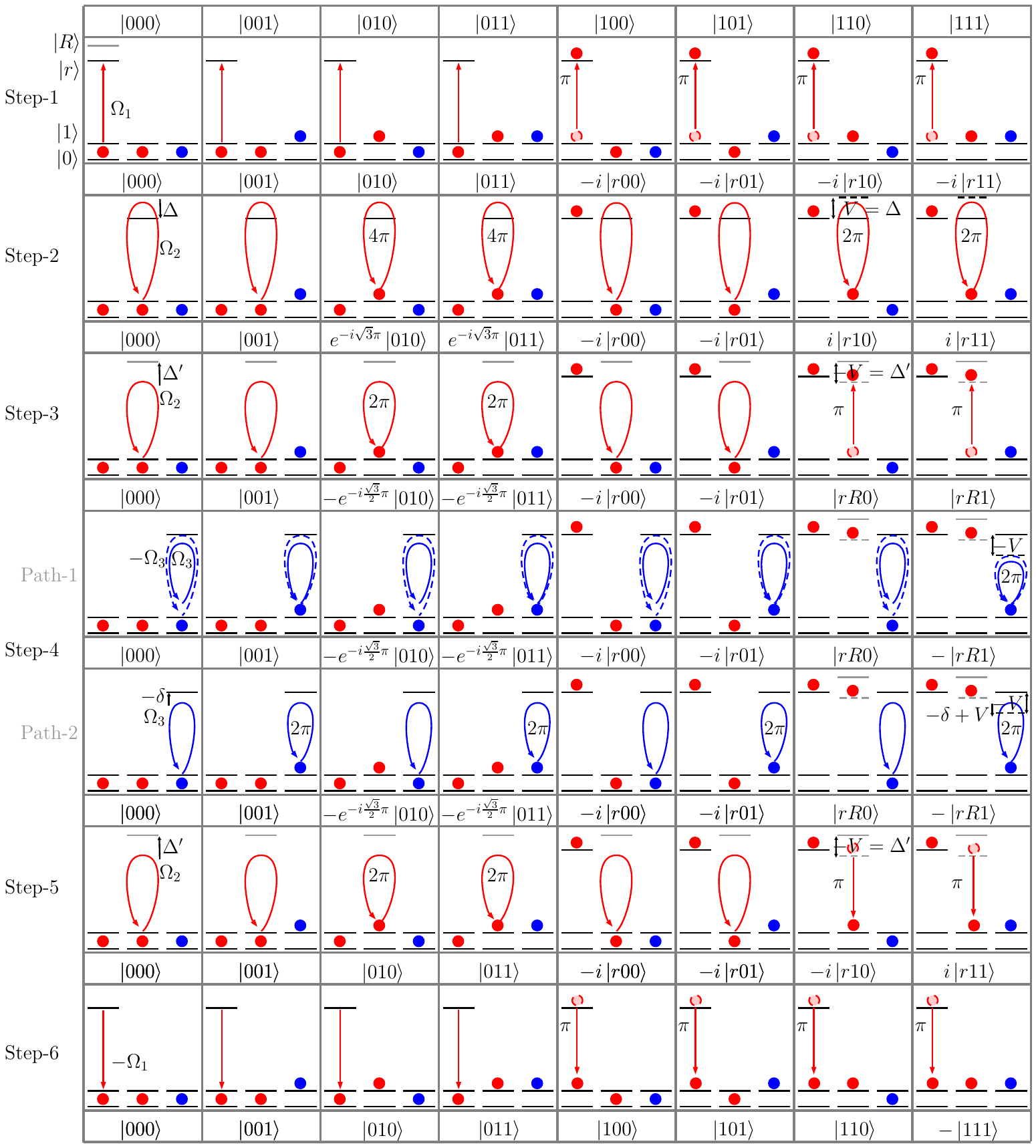}
    \caption{Schematic representation of the six-step CCZ-gate protocol. Red spheres and arrows denote the control atoms and applied lasers on them, whereas blue spheres and arrows are for the target atom. The qubit states are $\ket{0}$ and $\ket{1}$, and the Rydberg states are $\ket{r}$ and $\ket{R}$. For further details, see the main text.}
    \label{fig2}
\end{figure*}

As illustrated in Fig. \ref{fig1}(a), we select an array of three neutral atoms trapped in equidistant optical tweezers. As seen in Fig. \ref{fig1}(b), all the atoms are two-level systems in which the excited state is the Rydberg state and the qubits ($\ket{0(1)}$) are ground-state hyperfine levels. Since each atom may be addressed separately, we use successive pulses to build the gate protocol.  For each of the three atoms, $\ket{0}$ is completely decoupled and $\ket{1}$ is connected to the Rydberg state. The six-step gate protocol is covered in the following, and Fig. \ref{fig2} provides a schematic representation.

\paragraph{Step-1}
We begin with the initial states 
\begin{align*}
    \{\ket{000},\ket{001},\ket{010}, \ket{011},\ket{100},\ket{101},\ket{110}, \ket{111}\}
\end{align*}
 and apply a resonant $\pi$ pulse on the first control atom with the Rabi frequency $\Omega_1$ to get
 \begin{align*}
      \{\ket{000},\ket{001},&\ket{010}, \ket{011},\\&-i\ket{r00},-i\ket{r01},-i\ket{r10}, -i\ket{r11}\}.
 \end{align*}
 \paragraph{Step-2}
 A pulse with detuning $\Delta=V_1=V$ is applied on the second control atom. The state $\ket{010},\ket{011}$ will face an off-resonance transition while $\ket{r10},\ket{r11}$ will go through a resonant transition as the vdW interaction $V_1$ between the Rydberg states $\ket{r}$ of the first two control atoms will nullify the detuning $\Delta$. The pulse has to be continued for a  time ($t_0$) so that we cam complete a $4\pi(2\pi)$ rotation for $\ket{010},\ket{011}(\ket{r10},\ket{r11})$. Satisfying this condition, we get
\begin{equation}
    \begin{split}
        t_2=2\pi/\Omega_2=&4\pi/\sqrt{|\Omega_2|^2+\Delta^2}\\
        \Delta=&\sqrt3 \Omega_2.
    \end{split}
\end{equation}
The resulting states are 
\begin{align*}
    \ket{000},\ket{001},&e^{-i\sqrt3 \pi}\ket{010}, e^{-i\sqrt3 \pi}\ket{011},\\&-i\ket{r00},-i\ket{r01},i\ket{r10}, i\ket{r11}\}.
\end{align*}
\paragraph{Step-3}
A detuned pulse is applied to couple $\ket{1}$ to another Rydberg state $\ket {R}$ of the second control atom. $\ket{R}$ is chosen so that we can have  a negative blockade shift $V_2=-V$ (for detail see Sec.{\ref{sec:results}}). We fix the detuning $\Delta'$ to be equal to $V_2$ so that the Rydberg interaction can be nullified. Now we will adjust the time duration $t_1$ of the pulse in such a way that we can complete a $2\pi(\pi)$ rotation for $\ket{010},\ket{011}(\ket{r10},\ket{r11})$. Satisfying this, we again get $|\Delta'|=\sqrt3 \Omega_2$. Finally, the resulting states are 
\begin{align*}
    \{\ket{000},\ket{001},&-e^{-i\sqrt3\pi /2}\ket{010}, -e^{-i\sqrt3 \pi/2}\ket{011},\\&-i\ket{r00},-i\ket{r01},\ket{rR0}, \ket{rR1}\}.
\end{align*}
\paragraph{Step-4} We will discuss two different paths to complete this step.\\
1) Path-1: We apply a pulse with Rabi frequency $\Omega_3$ on the target atom for 
\begin{equation}
    t_4/2=2\pi/\sqrt{|\Omega_3|^2+V^2}.
\end{equation}
When the second control atom is at $\ket{R}$ the Rydberg state $\ket{r}$ of the target atom will face a shift $-V$. This results in a complete $2\pi$ pulse on the target atom with a phase 
\begin{equation}
    \theta/2=-\pi(1-V/\sqrt{|\Omega_3|^2+V^2}).
\end{equation}
But for the states where there is no vdW interaction between the second control atom and the target atom the $2\pi$ rotation remains incomplete. Now we apply another pulse for the time $t_4/2$ with the Rabi frequency $-\Omega_3$ so that $\ket{rR1}$ acquire a phase of
\begin{equation}
    \theta=2\pi V/\sqrt{|\Omega_3|^2+V}.
\end{equation}
Rests of the states will face an effective identity operation on them.
For $\Omega_3=\sqrt3V$, $\theta$ becomes $\pi$ and we get
\begin{align*}
    \{\ket{000},\ket{001},&-e^{-i\sqrt3 /2\pi}\ket{010}, -e^{-i\sqrt3 /2\pi}\ket{011},\\&-i\ket{r00},-i\ket{r01},\ket{rR0},-\ket{rR1}\}.
\end{align*}\\
2) Path-2: We apply a similar pulse like path-1 with the Rabi frequency $\Omega_3$, but with a detuning $-\delta$. When the second control atom is at $\ket R$ the effective detuning becomes $-\delta+V$ as the state $\ket r$ of the target atom gets shifted by $-V$ due to the RRI. We tailor the pulse applied to the target atom such that it produces a full $2\pi$ rotation in both cases, irrespective of whether RRI is present or absent. We choose $\delta=V/2$ so that the effective detuning becomes $-V/2$ and $V/2$ and the time taken to complete the pulse is
\begin{equation}
    t_4'=2\pi/\sqrt{|\Omega|^2+V^2/4}.
\end{equation}
The phase acquired by the target qubit when the control atom is (not) excited to $\ket{R}$ are
\begin{equation}
    \begin{split}
        \theta_V&=\pi+\pi(V-\delta)/\sqrt{|\Omega^2|+(V-\delta)^2},\\
        \theta&=\pi+-\pi\delta/\sqrt{|\Omega^2|+\delta^2}.
    \end{split}
\end{equation}
For $\Omega=\sqrt3V/2$ and $t_4'=2\pi/V$ the phases become
\begin{equation}
    \begin{split}
        \theta_V&=3\pi/2,\\
        \theta&=\pi/2.
    \end{split}
\end{equation}
This results in 
\begin{align*}
    \{\ket{000},&e^{-i\pi/2}\ket{001},-e^{-i\sqrt3\pi /2}\ket{010},\\& -e^{-i\sqrt3 \pi/2}e^{-i\pi/2}\ket{011},-i\ket{r00},\\&-ie^{-i\pi/2}\ket{r01},\ket{rR0}, -e^{-i\pi/2}\ket{rR1}\}.
\end{align*}
After giving a single qubit rotation, $\mathbb{I}_2\otimes\mathbb{I}_2\otimes R_Z(\pi/2)$ we get
\begin{align*}
    e^{-i\pi/4}\{\ket{000},&\ket{001},-e^{-i\sqrt3\pi /2}\ket{010}, -e^{-i\sqrt3 \pi/2}\ket{011},\\&-i\ket{r00},-i\ket{r01},\ket{rR0}, -\ket{rR1}\}.
\end{align*}

In this step, if the first control atom is at $\ket r$ and the second control atom is at $\ket R$, they will accumulate a phase $-Vt_4(-Vt_4')$ for path-1(2) because of RRI between them. We can disregard this since the accumulated phase for path-1(2) is $-2\pi$ since $t_4=t_4'=2\pi/V$..
\paragraph{Step-5} We repeat the third step to get
\begin{align*}
    \{\ket{000},\ket{001},&\ket{010}, \ket{011},\\&-i\ket{r00},-i\ket{r01},-i\ket{r10}, i\ket{r11}\}.
\end{align*}
\paragraph{Step-6:}
 We repeat the first step with $-\Omega_1$ to get the complete gate operation of the CCZ gate $\diag{(1,1,1,1,1,1,1,-1)}$.
 For the path-2 we get the esteemed operator with a common phase factor $\pi/4$.

\section{Gate Error Analysis}
\label{err}
The dominant contribution to the infidelity of the implemented CCZ gate arises from spontaneous decay of the Rydberg states. To quantify the effect of Rydberg decay, we define an associated error parameter, $\varepsilon = T_R/\tau$, where $T_R$ is the total time that the atoms collectively occupy their respective Rydberg states during the gate operation, and $\tau$ is the intrinsic radiative lifetime of those Rydberg states. 

Let us consider our CCZ gate where the two control and one target atom can be either in $\ket{0}$ or $\ket{1}$ initially. 
 When the atoms are in state $\ket {1}$, they can be excited to $\ket{R}$ with Rabi frequency $\Omega$ and detuning $\Delta$, and that atom evolves following the Hamiltonian
\begin{equation}
    H=-\frac{\Omega}{2}\ket{R}\bra{1}+ \Delta\ket R\bra R +\text{H.c.}, 
\end{equation}
which has the following two eigenstates:  
 \begin{equation}
     \begin{split}
         \ket{+}&=\cos\phi \ket R+\sin \phi\ket{1}\\
    \ket{-}&=\cos \phi\ket{1}-\sin\phi \ket R,
     \end{split}
 \end{equation}
 where 
  \begin{equation}
   \begin{split}
       \cos\phi&=\sqrt{1/2(1+ \Delta/\Omega_{\rm eff})}  \\ 
  \sin\phi&=\sqrt{1/2(1- \Delta/\Omega_{\rm eff})},
   \end{split}   
  \end{equation} 
 with the corresponding eigenvalues are 
 \begin{equation}
     \lambda_{\pm}=1/2( \Delta\pm\Omega_{\rm eff}),
 \end{equation}
 where $\Omega_{\rm eff}=\sqrt{|\Omega|^2+\Delta^2}$. The time evolution operator $U=e^{-iHt}$ acts on $\ket 1$ to give
\begin{equation}
    \begin{split}
      e^{-iHt}  \ket 1&=\sin\phi e^{-i\lambda_+t}\ket{+}+\cos\phi e^{-i\lambda_-t}\ket{-}\\
      &=e^{- i\Delta t/2}(\sin\phi e^{-i\Omega_{\rm eff}t/2}\ket{+}+\cos\phi e^{+i\Omega_{\rm eff}t/2}\ket{-}).
    \end{split}
\end{equation}

The Rydberg state $\ket R$ can be expressed as
\begin{equation}
    \ket R =\cos \phi \ket + -\sin \phi \ket -.
\end{equation}
The overlap of the state $\ket R$ in the time-evolved $\ket{1}$ state, given by $\bra{R}e^{-iHt}  \ket 1$, can be expressed as
\begin{equation}
\begin{split}
     e^{\mp i\Delta t/2}(\sin \phi \cos \phi e^{-\Omega_{\rm eff}t/2}-\sin \phi \cos \phi e^{\Omega_{\rm eff}t/2})\\=-ie^{\mp i\Delta t/2}\sin2\phi \sin \frac{\Omega_{\rm eff}t}{2}
\end{split}
\end{equation}
The time integral of the population in the state $\ket R$ during pulse duration $t_i$ can be calculated as 
\begin{equation}
\label{timeint}
    \begin{split}
        T_R&=\int_0^{t_i}\sin^2 2\phi \sin^2 \frac{\Omega_{\rm eff}t}{2}dt\\
        &=\frac{\Omega^2}{\Omega^2 + \Delta^2}\int_0^{t_i}\sin^2 \frac{\Omega_{\rm eff}t}{2}dt.
    \end{split}
\end{equation}

For zero detuning ($\Delta =0$), Eq.~\ref{timeint} reduces to 
\begin{equation}
\label{timeint2}
        T_R=\int_0^{t_i}\sin^2 \frac{\Omega t}{2}dt.
\end{equation}

\subsection*{ Path 1} We analytically compute the time integral  of the population of the Rydberg states for all control and target atoms for all eight computational basis states as follows:
\begin{equation}
    \begin{split}
    T_{000}&=0;\\
    T_{001}&=\big(\sqrt3\pi - sin(\sqrt3\pi)\big)/\Omega_3;\\
    T_{010}&=\pi/2\Omega_2;\\
    T_{011}&=\pi/2\Omega_2 + 3\sqrt3\pi/4\Omega_3;\\
    T_{100}&= \pi/\Omega_1 + 4\pi/\Omega_2 + 2\sqrt3\pi/\Omega_3 ;\\
    T_{101}&=\pi/\Omega_1 + 4\pi/\Omega_2 + \big(3\sqrt3\pi - sin(\sqrt3\pi)\big)/\Omega_3;\\
    T_{110}&=\pi/\Omega_1 + 6\pi/\Omega_2 + 4\sqrt3\pi/\Omega_3;\\
    T_{111}&=\pi/\Omega_1 + 6\pi/\Omega_2 + 19\sqrt3\pi/4\Omega_3;\\
    \end{split}
\end{equation}
For $\Omega_2 = V/\sqrt3$ and $\Omega_3 = \sqrt3V$, the time-integrated Rydberg state population, averaged over all computational basis states, for the CCZ gate in Path 1 is obtained as
\begin{equation}
    T_{R_{P1}} = \frac{\pi}{2\Omega_1} + \big(\frac{21\sqrt{3}}{8}\pi + \frac{31}{16}\pi - \frac{1}{4\sqrt{3}}\sin(\sqrt{3}\pi)\big)/V
\end{equation}

\subsection*{ Path 2} Similarly, we also calculate the time integral  of the population of the Rydberg states for all control and target atoms for all eight computational basis states in Path 2 as follows:
\begin{equation}
    \begin{split}
    T_{000}&=0;\\
    T_{001}&=3\sqrt3\pi/8\Omega_3;\\
    T_{010}&=\pi/2\Omega_2;\\
    T_{011}&=\pi/2\Omega_2 + 3\sqrt3\pi/8\Omega_3;\\
    T_{100}&= \pi/\Omega_1 + 4\pi/\Omega_2 + \sqrt3\pi/\Omega_3 ;\\
    T_{101}&=\pi/\Omega_1 + 4\pi/\Omega_2 + 11\sqrt3\pi/8\Omega_3;\\
    T_{110}&=\pi/\Omega_1 + 6\pi/\Omega_2 + 2\sqrt3\pi/\Omega_3;\\
    T_{111}&=\pi/\Omega_1 + 6\pi/\Omega_2 + 19\sqrt3\pi/8\Omega_3;\\
    \end{split}
\end{equation}
For $\Omega_2 = V/\sqrt3$ and $\Omega_3 = \frac{\sqrt3}{2}V$, the time-integrated Rydberg state population, averaged over all computational basis states, for the CCZ gate in Path 2 is obtained as
\begin{equation}
    T_{R_{P2}} = \frac{\pi}{2\Omega_1} +(21\sqrt{3}/8 + 15/8)\pi/V,
\end{equation}

\begin{figure}
    \centering
    \includegraphics[width=0.67\linewidth]{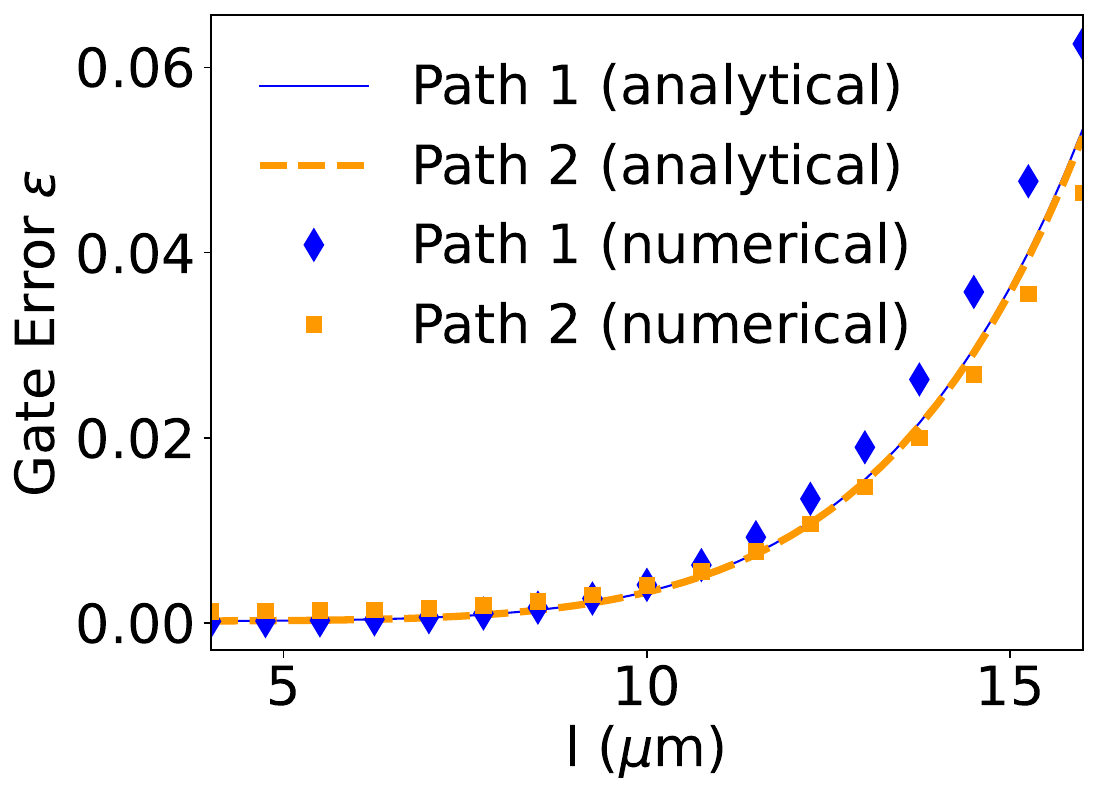}
    \caption{ Analytically calculated gate error due to the Rydberg decay for both paths is plotted by varying interatomic distance $l$ and also compared with the numerical results. We fix $\Omega_1/2\pi=1$MHz during this computation.}
    \label{error}
\end{figure}

In Fig.~\ref{error}, we show how $\varepsilon$ varies with interatomic separation for both paths, while we fix $\Omega_1/2\pi$ = 1 MHz. We compare these analytical results with the ones obtained numerically $(1-\bar{F})$, where $\bar{F}$ is the average gate fidelity, obtained by solving trace-preserving-operator (TPO) expression explained in the Sec. \ref{sec:results}.


\section{Results and Discussions}
\label{sec:results}
\begin{figure*}
    \centering
    \includegraphics[width=0.32\linewidth]{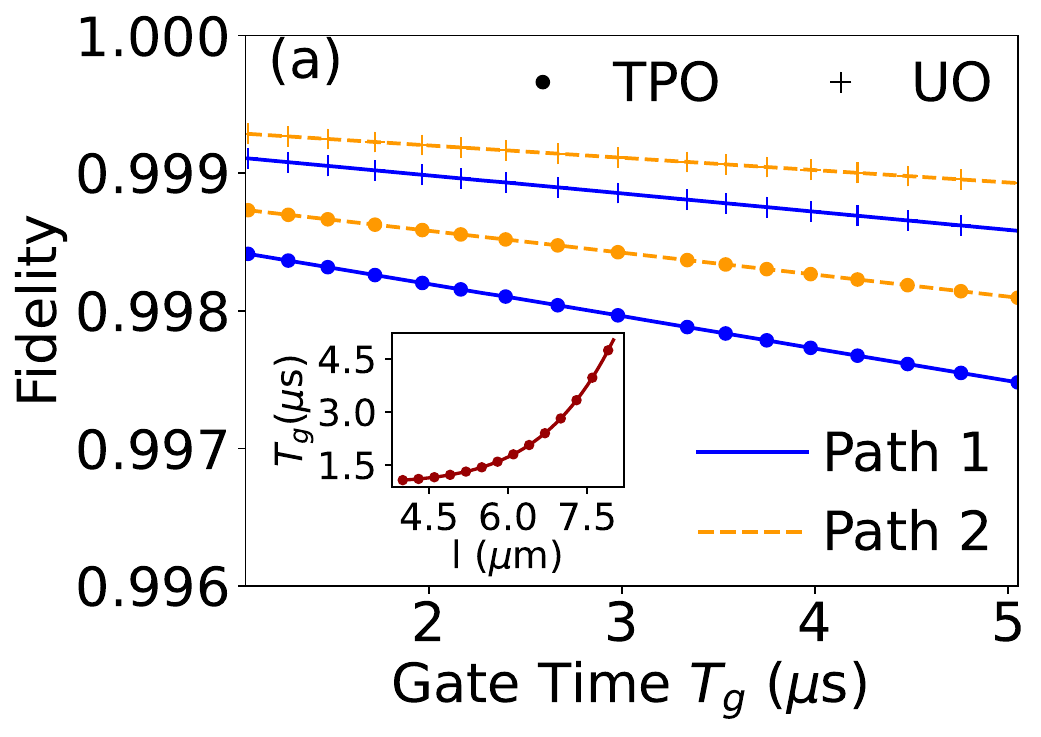}
    \includegraphics[width=0.32\linewidth]{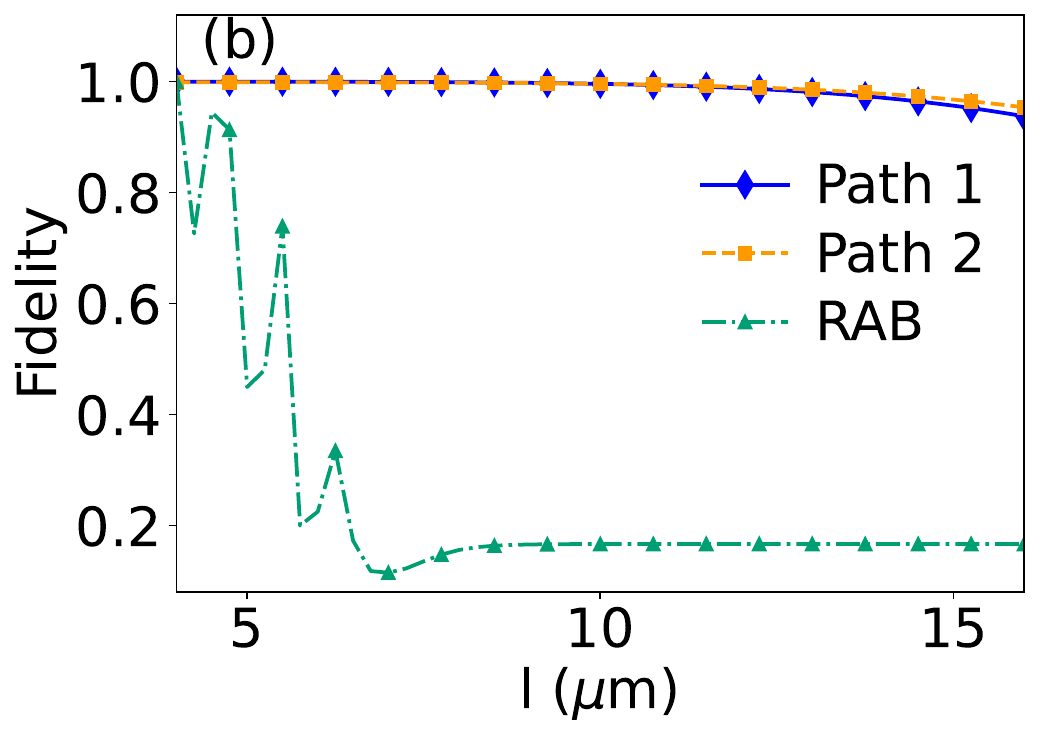}
    \includegraphics[width=0.32\linewidth]{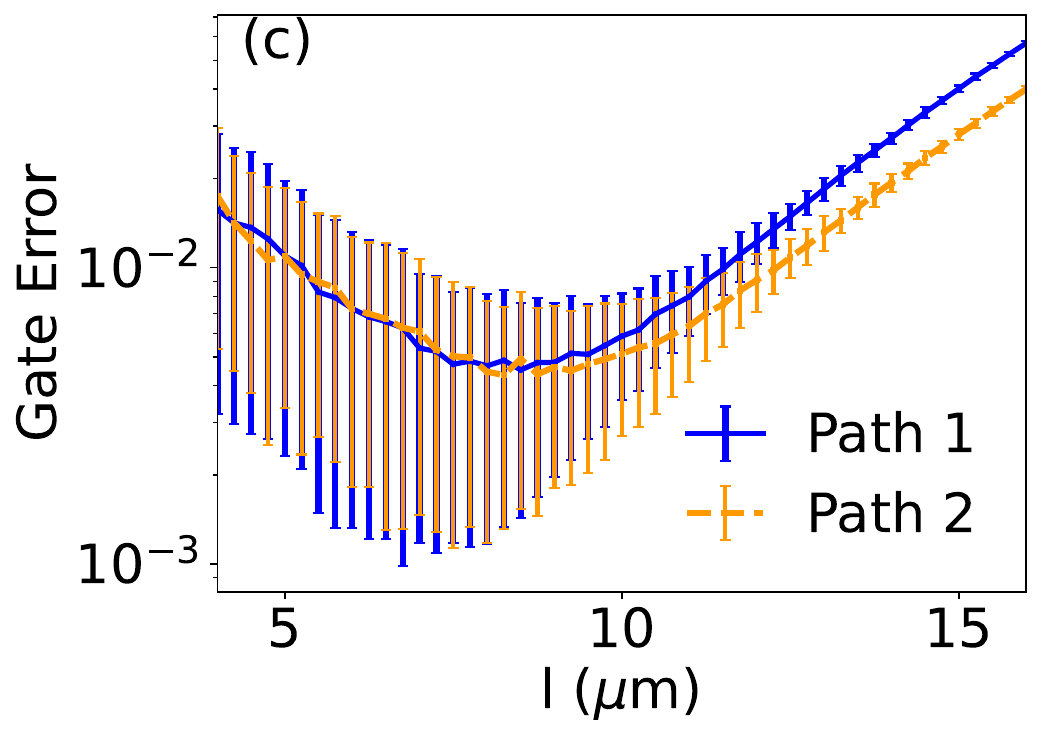}
    \caption{
(a) Gate fidelity versus total gate time $t_g$ for both proposed paths, evaluated using the TPO and UO measures. The inset shows $t_g$ as a function of the interatomic separation $l$.
(b) Comparison with the CCZ gate based on RAB, showing higher fidelities for the proposed paths at larger separations.
(c) Gate error $1-\bar{F}$ versus $l$ for both paths, obtained from Monte Carlo simulations incorporating thermally induced fluctuations in the atomic positions.
The trapping parameters are $\lambda_f=1.064~\mu\mathrm{m}$, $w_{f0}=1~\mu\mathrm{m}$, $T_a=2~\mu\mathrm{K}$, and $U_m=1~\mathrm{mK}$, with the trap depth expressed in temperature units. The Rabi frequency is fixed at $\Omega_1/2\pi=1~\mathrm{MHz}$ throughout.
}
    \label{fig:Fig 3}
\end{figure*}

\newcommand{\XX}[1]{\textcolor{red}{[#1]}}   

To illustrate the protocol numerically, we consider $^{87}\mathrm{Rb}$ atoms. The qubit states $\ket{0}$ and $\ket{1}$ are encoded in selected Zeeman sublevels of the $5S_{1/2}(F=1)$ and $5S_{1/2}(F=2)$ ground-state hyperfine manifolds, respectively. We choose $\ket{r}=\ket{80P_{3/2},m_J=3/2}$ and $\ket{R}=\ket{90P_{3/2},m_J=3/2}$ as the two Rydberg states. For these Rydberg levels, we take $m_J=3/2$ and an internuclear axis
parallel to the quantization axis. The long-range interactions originate
from virtual electric dipole transitions to nearby pair states, yielding
second-order vdW shifts~\cite{Walker2008,SaffmanWalkerMolmer2010}.
Conservation of the total magnetic projection $M=3$ restricts the
intermediate configurations to $D+D$ and $D_{5/2}+S_{1/2}$ pairs,
including both atomic orderings; $S+S$ and $D_{3/2}+S_{1/2}$ channels
do not contribute in this geometry. The resulting pair-state shifts are
\begin{equation}
 V_{rr}(l)=\frac{C_6^{rr}}{l^6},\qquad
 V_{rR}(l)=\frac{C_6^{rR}}{l^6},
 \label{eq:pair_vdw_results}
\end{equation}
where $\hbar V_{rr}$ and $\hbar V_{rR}$ are interaction energies.
For the selected $^{87}\mathrm{Rb}$ states, Appendix C of
Ref.~\cite{Shi2017BlockadeFree} reports $C_6^{rr}/2\pi=289$ and
$C_6^{rR}/2\pi=-281$, in units of
$\mathrm{GHz}\,\mu\mathrm{m}^{6}$. These coefficients give repulsive
and attractive shifts of similar magnitudes, allowing the sign of the
conditional detuning to be changed by selecting the Rydberg level.
The quoted crossover distances are approximately $0.74$ and
$1.3~\mu\mathrm{m}$, respectively; a controlled van der Waals
description requires separations sufficiently beyond the larger scale.

The mixed pair also permits second-order exchange
$\ket{rR}\leftrightarrow\ket{Rr}$, which appears as an off-diagonal
matrix element in the effective pair Hamiltonian~\cite{Shi2014}.
For the present $80P_{3/2}$--$90P_{3/2}$ combination, the reported
exchange coefficient is approximately four orders of magnitude smaller
than the diagonal coefficient~\cite{Shi2017BlockadeFree}. Exchange may therefore
be neglected provided $|J_6|t_g/l^6\ll1$, where $J_6/l^6$ is the
exchange coupling in angular-frequency units and $t_g$ is the gate
duration.

Direct single-photon excitation from an alkali-metal $S$ ground state to
a Rydberg $P$ state is electric-dipole allowed and avoids scattering
through an intermediate optical level. Its experimental feasibility was
demonstrated by Hankin \textit{et al.}~\cite{Hankin2014}, who coherently
excited $^{133}\mathrm{Cs}$ on the
$6S_{1/2}\rightarrow84P_{3/2}$ transition using continuous-wave light
near $319$~nm and observed two-atom blockade at a separation of
$6.6(3)~\mu\mathrm{m}$. For the present rubidium implementation, the
corresponding transitions originate from $5S_{1/2}$; their wavelengths
and dipole matrix elements must be evaluated for $^{87}\mathrm{Rb}$.
Laser-frequency stability, electric-field compensation, and motional
control remain necessary to suppress dephasing.

We analyze the gate performance by computing the gate fidelity against different computational parameters.
The average gate fidelity is evaluated using the trace-preserving-operator (TPO) expression \cite{Nielsen2002-simple}
\begin{equation}\label{Efid}
   \bar{F}(\hat{\mathcal{O}},\epsilon)
   = \frac{\sum_j \mathrm{Tr}\!\left[\hat{\mathcal{O}}\,\hat{\mathcal{O}}_j^\dagger\,\hat{\mathcal{O}}^\dagger\,\epsilon(\hat{\mathcal{O}}_j)\right] + d^2}{d^2(d+1)} ,
\end{equation}
where $\{\hat{\mathcal{O}}_j\}$ is the complete set of tensor products of $N$-qubit Pauli operators, e.g.\ $\{\hat I\otimes\hat I\otimes\hat I,\ \hat I\otimes\hat I\otimes\hat\sigma_x,\ \ldots,\ \hat\sigma_z\otimes\hat\sigma_z\otimes\hat\sigma_z\}$ for three qubits. Here $\hat{\mathcal O}=U_{\rm CCZ}$ is the ideal gate, $\epsilon$ is the trace-preserving quantum operation realized by the physical implementation, and $d=2^N=8$. An alternative measure is the unitary overlap (UO),
\begin{equation}\label{Ufid}
    F=\frac{1}{d}\,\Big|\mathrm{Tr}\big[\hat U_{\rm ideal}^\dagger\,\hat U_{\rm phys}\big]\Big| ,
\end{equation}
where $\hat U_{\rm ideal}=U_{\rm CCZ}$ and $\hat U_{\rm phys}$ is the gate matrix obtained from the simulation, projected onto the computational subspace. In the presence of spontaneous emission from the Rydberg levels, the evolution of the whole system is governed by the master equation
\begin{equation}
    \dot{\hat{\rho}}=-i[\hat H, \hat \rho]+ \sum_{k=1}^{3}\sum_{i=0}^{1}\mathscr{D}\big[\sigma^{r}_{i,k}\big]\hat{\rho}+\sum_{i=0}^{1}\mathscr{D}\big[\sigma^{R}_{i,2}\big]\hat{\rho},
\end{equation}
where $\mathscr D[\hat a]\hat\rho=\hat a\hat\rho\hat a^\dagger-(\hat\rho\hat a^\dagger\hat a+\hat a^\dagger\hat a\hat\rho)/2$ and $\sigma^{\nu}_{i,k}=\sqrt{\gamma_\nu/2}\,\ket{i}_k\bra{\nu}$ describes decay of the $k$-th atom from the Rydberg level $\nu$ to the qubit state $\ket{i}$, with $\gamma_\nu=1/\tau_\nu$. 


Although Eqs.~\eqref{Efid} and \eqref{Ufid} use gate matrices obtained from the same master-equation simulation, they characterize different aspects of the gate. Eq.~\eqref{Efid} is the average gate fidelity \cite{Nielsen2002-simple}, a basis-averaged figure of merit for the completely positive trace-preserving map generated by the open-system evolution. It is the primary metric for the realistic implementation, because it captures coherent control errors, spontaneous emission and population leakage out of the computational subspace. By contrast, Eq.~\eqref{Ufid} measures the similarity between the effective gate obtained from the master equation and the ideal unitary within the computational subspace. It is a useful complementary benchmark of how closely the dynamics reproduce the desired gate structure, but it does not fully account for leakage and non-unitary errors. Further discussion of the fidelity metrics can be found in \cite{Kazemi2025-fg}.

In the Fig. \ref{fig:Fig 3}(a) we have shown how the gate fidelity varies with changing the total gate time ($t_g$). Here $t_g$ is defined by 
\begin{equation}
    t_g=\sum_{i=1}^{6}t_i,
\end{equation}
where $t_i$ is the time taken to complete $i$-th step. By simple calculation we get
\begin{equation}\label{Etimes}
t_1=t_6=\frac{\pi}{\Omega_1},\
t_2=\frac{2\sqrt3\pi}{V},\
t_3=t_5=\frac{\pi\sqrt3}{V},\
t_4=\frac{2\pi}{V}.
\end{equation}
The total gate time for both paths is as follows:
\begin{equation}
    t_g=2\pi/\Omega_1+4\sqrt{3}\pi/V+2\pi/V.
\end{equation}
The inset shows the change in $t_g$ while varying $l$.

In Fig.~\ref{fig:Fig 3}b, we compare the gate fidelity by using the TPO-based formalism of our proposed CCZ following Path 1 and Path 2 with the previous CCZ gate  (\cite{Su2017}) that works based on RAB mechanism. Since our gate works beyond strong blockade regimes, where the RAB-based CCZ fails, they clearly outperform the latter for large interatomic distances, as indicated by the results, thus allowing for long-distance gate operation.

Now, we model the position fluctuations of trapped atoms using the thermal widths of a Gaussian optical tweezer potential taking into account the effect of thermal atomic motion. For a trap wavelength $\lambda_f = 1.064~\mu\mathrm{m}$, trap waist $w_{f0}=1~\mu\mathrm{m}$, atomic temperature $T_a = 2~\mu\mathrm{K}$, and trap depth $U_m = 1\mathrm{mK}$, the time-averaged variance of the position spreads are obtained from Eqs (6a) and (6b) of \cite{SaffmanWalker2005}:
\begin{equation}
\braket{\sigma_x^2}=\braket{\sigma_y^2}=
{
\frac{w_{f0}^{\,2}  T_a}{4U_m}
},
\end{equation}
and
\begin{equation}
\braket{\sigma_z^2}=
{
\frac{\pi^2 w_{f0}^{\,4}  T_a}
{2\lambda_f^{\,2}U_m}
}.
\end{equation}
Substituting the value of the experimental parameters, we obtain
\begin{equation}
\sigma_x=\sigma_y \approx 0.02~\mu\mathrm{m},
\qquad
\sigma_z \approx 0.09~\mu\mathrm{m}.
\end{equation}
We incorporate these thermal fluctuations into the Monte Carlo simulations by randomly sampling (1000 shots) the atomic positions from Gaussian distributions with the above standard deviations. In Fig.~\ref{fig:Fig 3}c using the TPO-based formalism, we show how the gate error ($1-\bar{F}$) varies  with changing the interatomic distance, while the colour bar signifies the corresponding standard deviation of that particular point.


\section{Conclusion}
\label{sec:conclusion}

We have proposed a six-step protocol for implementing a CCZ gate in a neutral-atom platform, wherein three individually addressable atoms are confined in a linear array of equally spaced optical tweezers, and the strength of the RRI is comparable to the relevant Rabi frequencies. The protocol uses two different Rydberg states to access interaction-induced energy shifts of opposite sign and integrates sequential excitation of Rydberg atoms using analytically derived pulse sequences.
The conditional $\pi$ phase is imprinted on the target atom either by a pair of
pulses with opposite Rabi amplitudes or by a single detuned pulse followed by a
local $R_z(\pi/2)$ correction. We have analytically calculated the decay error by time-integrating Rydberg populations, and they agree with the infidelity obtained from master-equation
simulations for $^{87}$Rb. Within the parameter regime studied, the protocol
maintains high fidelity at separations where the fast antiblockade CCZ gate fails, and Monte Carlo sampling of thermal position
fluctuations quantifies its sensitivity to variations of the interaction
strength. These results identify the trade-offs among
gate duration, Rydberg-state exposure, and interaction sensitivity and provide an
analytically specified route to native three-qubit phase gates beyond the
strong-blockade regime.

\begin{acknowledgments}

The authors acknowledge the use of GPT-6 Astra (OpenAI) for assistance with language editing and improving the clarity and organization of the manuscript. The authors take full responsibility for the final manuscript, including its scientific content and conclusions.

\end{acknowledgments}


\appendix


\bibliography{CCZ}

\end{document}